\documentclass{mn2e} \usepackage{graphicx} \usepackage{epsfig}
\usepackage{float} 
\usepackage{color} \usepackage{ulem}

\def\starlight{\textsc{starlight}}                    
\newcommand{\hii}{\ifmmode [\rm{H}\,\textsc{ii}] \else [H~{\sc ii}]\fi}
\newcommand{\Ha}{\ifmmode {\rm H}\alpha \else H$\alpha$\fi}
\newcommand{\Hb}{\ifmmode {\rm H}\beta \else H$\beta$\fi}
\newcommand{\oiii}{\ifmmode [\rm{O}\,\textsc{iii}] \else [O~{\sc iii}]\fi}
\newcommand{\oii}{\ifmmode [\rm{O}\,\textsc{ii}] \else [O~{\sc ii}]\fi}
\newcommand{\oi}{\ifmmode [\rm{O}\,\textsc{i}] \else [O~{\sc i}]\fi}
\newcommand{\nii}{\ifmmode [\rm{N}\,\textsc{ii}] \else [N~{\sc ii}]\fi}
\newcommand{\sii}{\ifmmode [\rm{S}\,\textsc{ii}] \else [S~{\sc ii}]\fi}

\title[CGCG~292$-$057]
      {CGCG~292$-$057 -- a radio galaxy with merger-modulated radio activity}
 
\author[Kozie\l]
{D. Kozie\l -Wierzbowska$^{1}$,
  M. Jamrozy$^{1}$, S. Zola$^{1,2}$, G. Stachowski$^{2}$ and A. Ku\'zmicz$^{1}$\\
  $^{1}$Astronomical Observatory, Jagiellonian University,
ul. Orla 171, PL-30244 Krakow, Poland\\
$^{2}$Mt. Suhora Observatory, Pedagogical University, ul. Podchorazych 2, PL-30084 Krakow, Poland\\
  }

\begin{document}

\maketitle

\begin{abstract} 
We announce the discovery of a unique combination of features in a radio source
identified with the merger galaxy CGCG~292$-$057. The radio galaxy both exhibits 
a highly complex, X-like structure and shows signs of recurrent 
activity in the form of double-double morphology.
The outer lobes of CGCG~292$-$057 are characterized by low radio power, 
$P_{1400MHz} \simeq 2 \times 10^{24}{\rm~W\,Hz^{-1}}$, placing this source 
below the FRII/FRI luminosity threshold, and are highly polarized (almost 
20~per~cent at 1400 MHz) as is typical of X-shaped radio sources. The host 
is a LINER-type galaxy with a relatively low black hole mass and double-peaked 
narrow emission lines. 

These features make this galaxy a primary target for studies of merger-triggered 
radio activity. 

\end{abstract}

\begin{keywords} galaxies: active - galaxies: individual: CGCG~292$-$057 - galaxies: nuclei - galaxies: structure - radio continuum: galaxies
\end{keywords}


\section{Introduction}
\label{sec:Introduction}

Faranoff \& Riley (1974) proposed a morphological classification of radio sources which distinguished two classes of objects known as Faranoff-Riley type I (FRI) and type II (FRII) radio galaxies. However, sources
exist whose morphology cannot be fully described by either of these types. These
include radio galaxies with two pairs 
of lobes emerging from the same radio core and aligned along a common axis. 
This group of objects, named `double-double radio galaxies' (DDRGs, Schoenmakers 
et al. 2000), constitute a rare class of extragalactic radio sources (about 
two dozen objects in total; for references see e.g. Saikia \& Jamrozy 2009). 
In addition, two radio galaxies with three pairs of coaxial lobes (so called `triple-double') have also been detected: J0929$+$4146 (Brocksopp et a. 2007) and J1409$-$0302 (Hota at al. 2011).

Another group of peculiar sources are radio galaxies with an `X-shaped' morphology 
(Leahy \& Parma 1992), in which the second pair of lobes (or `wings') 
is oriented along a substantially different axis, leading to the formation of an X-like structure.
These wings are symmetrical about the radio core, in some cases exhibiting a Z-shaped 
symmetry (Gopal-Krishna et al. 2003). They may be long, sometimes even 
longer than the primary lobes (e.g. 3C223.1 and 3C403, Dennett-Thorpe et al. 2002; 
4C+00.58, Hodges-Kluck et al. 2010; 4C+32.25, Parma, Ekers \& Fanti 1985).

Several authors have attempted to explain the formation of X-shaped structures and there exist a number of models, 
interpreting the morphology as the result of: (i) hydrodynamic backflow (Leahy \& Williams 1984; Worrall, Birkinshaw \& Cameron 
1995; Hodges-Kluck \& Reynolds 2011), (ii) conical precession (Parma, Ekers \& Fanti 1985), (iii) fast 
realignment of the jet (e.g. Dennett-Thorpe et al. 2002; Merritt \& Ekers 2002), (iv) existence 
of an unresolved binary AGN system with two pairs of jets (Lal \& Rao 2007), or (v) interaction of 
jet with merger remnants within the host galaxy (Gopal-Krishna et al. 2012). A comprehensive review 
of the existing models is given by Gopal-Krishna et al. (2012). Our knowledge of these peculiar 
sources is growing continuously, and the number of such sources known is increasing -- for example, candidate 
X-shaped sources were found by Cheung (2007).

The fact that in some radio galaxies two pairs of lobes are observed implies that the time required for the 
jet flow to cease is shorter than that for the outer lobes
to fade. Parma et al. (2007) showed that the typical age of the active phase 
(derived from spectral fitting) is $10^7$--$10^8$ years and that the relic phase may be 
comparable in its duration. Therefore, since the lobes of extended radio 
sources can store the energy supplied by the jets for a time longer than the 
duration of the quiescent  active galactic nucleus (AGN) phase, radio galaxies are able to preserve 
information on the past activity of the AGN. Hence, both DDRG and X-shaped sources
are believed to be signatures of intermittent AGN activity, although reflecting distinct physical mechanisms. 

The growth of the central black hole and the activity of the AGN are 
regulated by the supply of material in the central regions of the galaxy. An increase 
in the amount of gas in the central region and a reorientation of the jet 
axis can be caused by interaction or merger with a neighbouring galaxy. 
The widely accepted idea is that such interactions generate flows of gas from the 
outer parts of a galaxy into its inner regions, through loss of angular momentum 
induced by tidal forces (Mihos \& Hernquist 1996). Signs of such interactions and mergers 
have been found in Seyferts (Keel et al. 1985), quasars (Sanchez \& Gonzalez-Serrano 2002; 
Heckman et al. 1984; Cattaneo et al. 2005a,b; Springel, Di Matteo \& Hernquist 2005) 
and radio galaxies (e.g. Wilson \& Colbert 1995).  
Unfortunately, as shown by Dennett-Thorpe et al. (2002), X-shaped radio sources have not
usually been identified with merger galaxies. The only exceptions known so far are 
3C305 (Heckman et al. 1982) and 3C293, 
which shows tidal tails and a close companion galaxy (Evans et al. 1999; Martel et al. 
1999). An example combining both the X-shaped and double-double 
radio structure is also known: 4C+12.03 is a well studied source (Merritt \& Ekers 2002; 
Capetti et al. 2002). 

In this paper we draw attention to the unique radio source associated with CGCG~292$-$057, 
which combines many of the features described above. It shows signs of recurrent 
activity, in the form of a double-double morphology and an X-like radio structure similar to 4C+12.03, 
and it is identified with a merger galaxy. This combination of features makes this galaxy one of a kind.

Throughout this paper we assume a $\Lambda$ Cold Dark Matter cosmology with 
$H_{0}=71$km\,s$^{-1}$Mpc$^{-1}$, $\Omega_{\rm m}=0.27$, and $\Omega _{\Lambda}=0.73$ 
(Spergel et al. 2003).


\section{CGCG~292-057 -- the host galaxy}
\label{sec:optical}

The radio galaxy is identified with the bright (R $\simeq$ 14mag) galaxy CGCG~292$-$057. 
This galaxy is located on the sky at RA = 11$^{\rm h} $59$^{\rm m} $05\fs7, Dec = +58\degr~20$^\prime$ 36\farcs (J2000.0) 
and lies at a redshift of $z=0.054$ (Stoughton et al. 2002). As in case of most radio sources, we might expect the host to be an 
elliptical galaxy, but the Sloan Digital Sky Survey (SDSS; York et al. 2000) images show a disc-dominated 
structure.
The host galaxy morphology can be roughly determined using the so-called `concentration index', 
defined as the ratio of radii containing 90~per~cent and 50~per~cent of the Petrosian $r$ light, 
C$ \equiv r_{90}/r_{50}$ (Strateva et al. 2001, Shimasaku et al. 2001). Elliptical galaxies 
are more centrally concentrated than the exponential discs of spiral galaxies and the value of 
$C=2.83$ separates early- and late-type galaxies with a reliability of 81~per~cent (Strateva et al. 2001). 
The concentration index of CGCG~292$-$057 is 2.81$\pm$0.03, which places this galaxy almost exactly 
at the separation line.

\subsection{Optical observations}
\label{sec:imaging}

Since the SDSS images of CGCG~292$-$057 were insufficient to unambiguously classify the 
morphology, we performed new optical observations of this galaxy. 

In order to image the external structure of CGCG~292-057, 
we used the 60-cm telescope at the Mt. Suhora Astronomical Observatory of the Cracow 
Pedagogical University. The telescope is equipped with an SBIG ST10XME CCD camera 
mounted at the prime focus and a set of broadband filters (Bessell 1990).
The observations were performed in the $V$, $R$ and $I$ filters during 4 nights between 
2010 January 24 and 2010 April 19. The total net integration time in each filter was 
about 4 hours, split into 1-3 minutes exposures.
Observations were reduced for  bias, dark current and flat-field using the \textsc{iraf} package. 
After removal of cosmic rays, all the images in each filter were averaged into one optical map.

\begin{figure}
\centering
\includegraphics[width=8cm,angle=0]{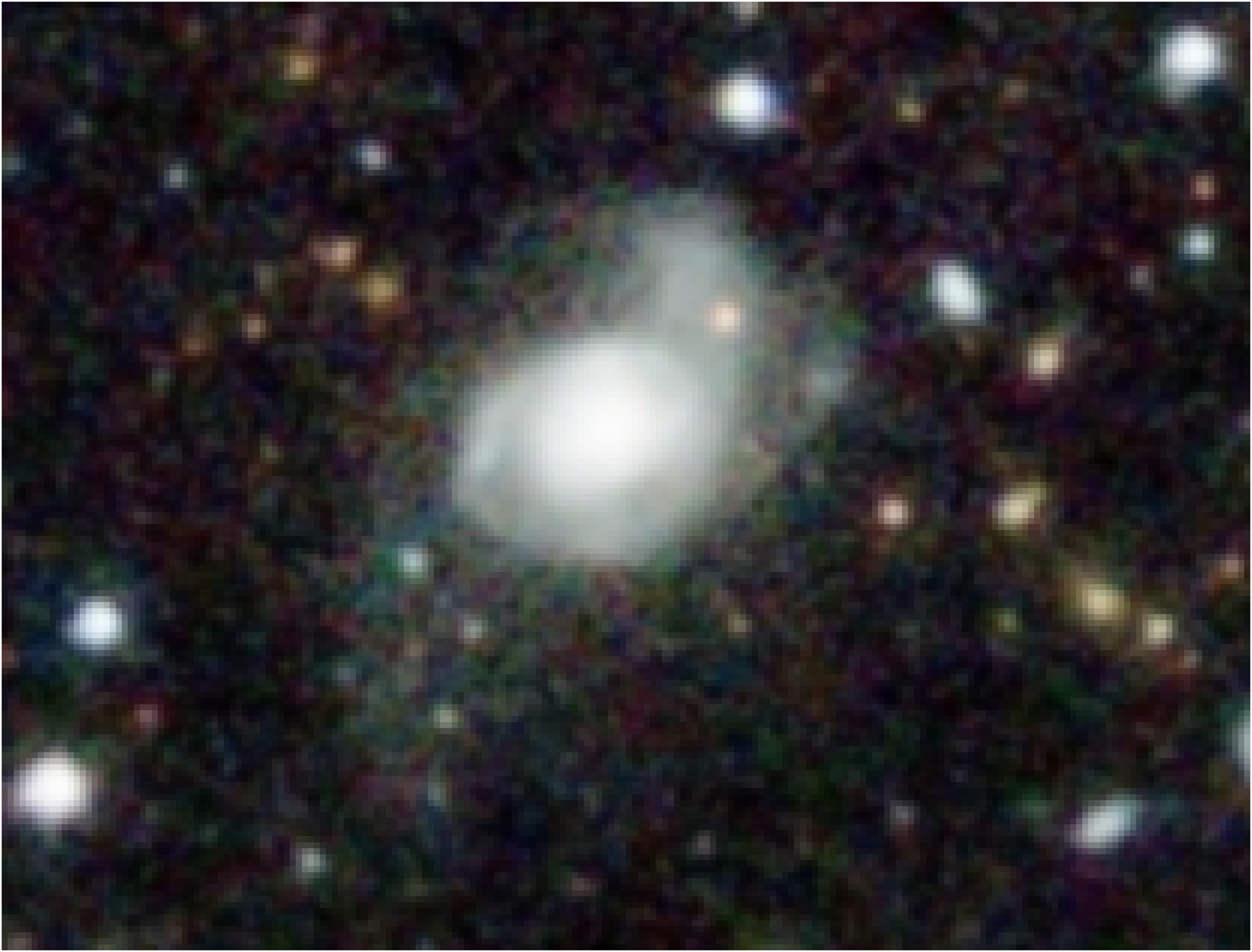}
\includegraphics[width=8cm,angle=0]{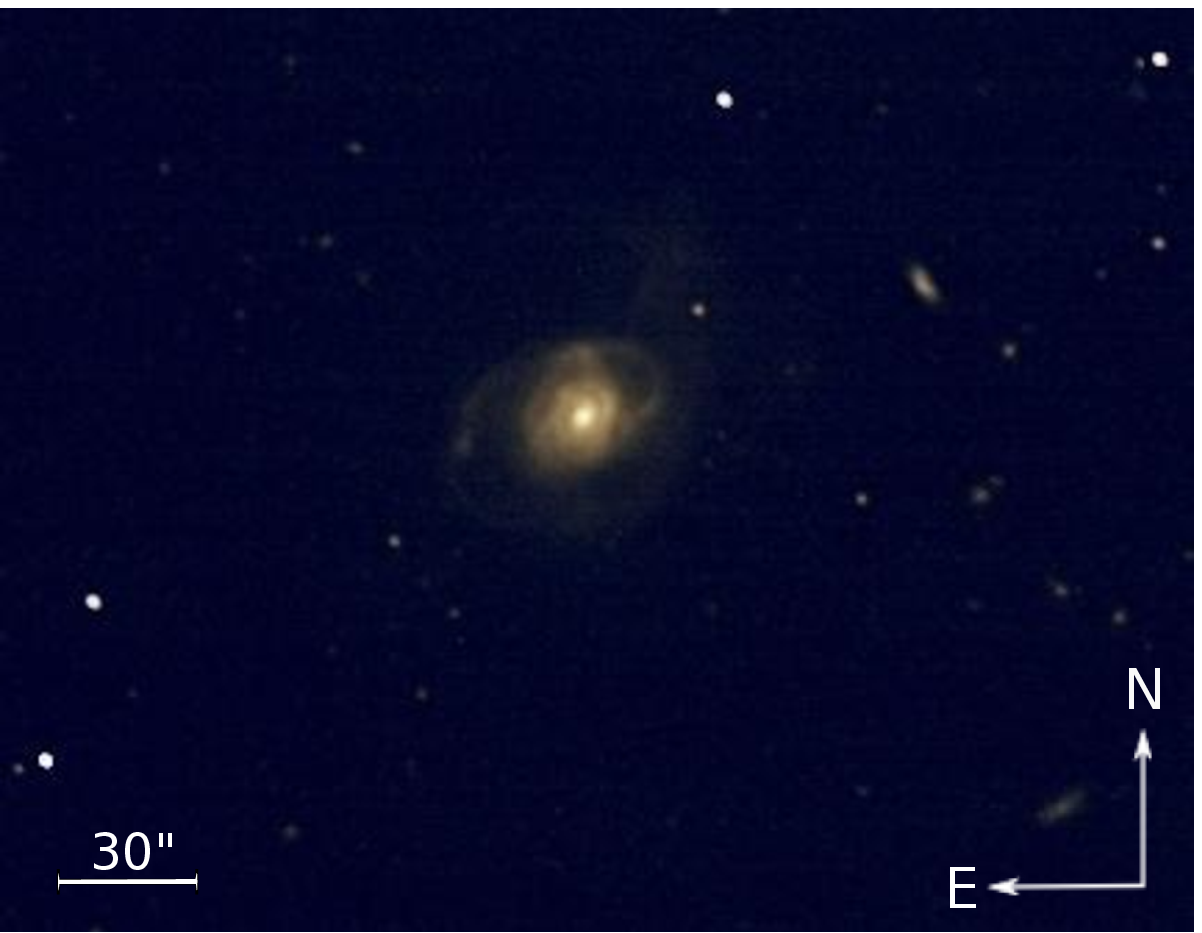}
\caption{Color images of CGCG~292$-$057 obtained with (top) the 60-cm telescope at Mt. Suhora and (bottom) the 2-m Faulkes Telescope. Both images have the same scale and orientation.}
\label{fig:imaging}
\end{figure}

An attempt to obtain images with better spatial resolution was undertaken with 
the 2-m Faulkes Telescope North operated by the Las Cumbres Observatory Global 
Telescope (LCOGT) Network. Observations were made in two instrumental configurations: 
in the Bessel $B$ filter using a 2k$\times$2k e2v chip CCD42-40 with an image scale of 0.2785~arcsec~pix$^{-1}$, 
and in the SDSS $r$ filter using a 4k$\times$4k Fairchild CCD486 BI with an image scale of 0.304~arcsec~pix$^{-1}$.
Our target galaxy was observed during 3 nights between 6 and 15 November 2010. 
Again, frames reduced for bias, dark current and flat-field corrections with 
the telescope pipeline were averaged to obtain images with a better signal-to-noise ratio.

The results of the optical observations combined into separate color images are shown 
in Fig.~\ref{fig:imaging}. The low spatial resolution but better sensitivity image from Mt. Suhora (Fig.~1, 
top panel) shows expanded, faint arms and diffuse, scattered emission from a distance of $\sim$1 
arcmin from the nucleus. The bottom panel in Fig.~1 shows the nucleus and the surroundings
with better resolution in the LCOGT image. Much of the fine structure observed in our deep LCOGT images 
is an indication of a merger event in the relatively recent past, probably within the 
last $\sim1$~Gyr. Our optical observations clearly show that the host galaxy CGCG~292$-$057 
cannot be considered to be an undisturbed spiral or, \textit{a fortiori}, an elliptical galaxy. 
The tidal debris forms structures resembling spiral arms, while the bright nucleus of this galaxy is 
surrounded by the tidal features and tails that only approximately recall a past spiral structure. 
CGCG~292$-$057 is observed almost `face-on' and undeniably is a merger galaxy. Unfortunately, 
none of the galaxies visible nearby can be considered to be the interacting companion.

\subsection{SDSS nuclear spectrum}
\label{sec:spectrum}

Spectral observations of CGCG~292$-$057 became available with the SDSS Early Data Release 
(Stoughton et al. 2002). The SDSS spectra were taken with a 3~arcsec fibre, which in 
the case of CGCG~292$-$057 covers 3.1~kpc of the central region, with a mean spectral 
resolution of $R=1800$ over a wavelength range from 3200~\AA\ to 9200~\AA. 

The spectrum of CGCG~292$-$057 has a bright stellar continuum with prominent absorption lines. 
The amplitude of the spectral discontinuity at the 4000~\AA\ break, D4000, strongly correlates 
with the mean age of galaxy stellar populations (Bruzual 1983, Cid Fernandes et al. 2005). D4000 
is equal to 1.83 for our target, which suggests evolved stellar populations in the centre of the
host galaxy (Bruzual 1983). 

\begin{figure}
\centering
\includegraphics[width=8cm,angle=0]{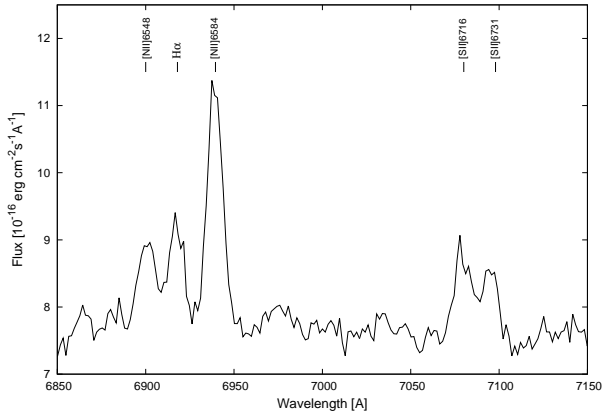}\caption{The nuclear SDSS spectrum of 
CGCG~292$-$057 in the wavelength range 6910--7130\AA\ in units of 
$10^{-16}$~erg\,cm$^{-2}$\,s$^{-1}$~\AA$^{-1}$. [NII], \Ha\ and [SII] show double-peak profiles.}
\label{fig:spectrum}
\end{figure}

Galaxies showing emission lines can be classified into star-forming and AGN hosting galaxies according to 
their positions on the diagnostic diagrams introduced by Baldwin, Phillips \& Terlevich (1981), while AGN 
hosts can be further classified as Seyferts or low-ionization nuclear emission-line region galaxies (LINERs)
using empirical dividing lines introduced by Kewley et al. (2006). 
A similar scheme for optical spectral classification of radio-loud galaxies was suggested by Laing et al. (1994),
 who proposed a separation into high- and low-excitation radio galaxies (HEG and LEG, respectively). This scheme
was revised by Buttiglione et al. (2010).

The four optical line ratios needed to classify CGCG~292$-$057 are given in Table 1. The emission line fluxes 
were taken from the \starlight\ database\footnote{http://www.starlight.ufsc.br} (Cid Fernandes et al. 2005; Mateus et al. 2006). 
These values indicate that our galaxy is an AGN host of the LINER type (Heckman 1980),
as well as an LEG according to the revised definition given by Buttiglione et al. (2010). 
In contrast to HEGs, which require large amounts of cold gas close to the AGN to produce the observed torus and the inferred thin, 
radiatively-efficient accretion disc, LEGs are supposedly powered by Bondi accretion of the hot phase of the intergalactic medium (IGM)
(e.g. Hardcastle et al. 2007; Hardcastle et al. 2009). The absence of strong, high excitation lines
in the spectrum of CGCG~292$-$057 indicates that this galaxy has a radiatively-inefficient active nucleus.

Fig.~\ref{fig:spectrum} shows the SDSS nuclear spectrum of CGCG~292$-$057 in the 
region of H$\alpha$ and [SII] lines. These lines also show peculiarities 
suggestive of double-peaked profiles. Double-peaked emission lines may be interpreted 
as indicators of a supermassive binary black hole (SMBBH) or the result of 
Narrow-Line Region kinematics (e.g. Liu et al. 2010). The fact that a companion galaxy is not visible
in the Mt. Suhora and LCOGT optical images would suggest that we are observing
the post-merger stage of the interaction, when the separation of the black holes is far below 
kpc scales and hence impossible to resolve spatially. If confirmed, the presence of double-peaked 
emission lines in the nuclear spectra may further indicate that the supermassive black holes 
from both galaxies have not yet merged. 

\begin{table}
\begin{center}
\begin{tabular}{l | l}
\hline
Line ratio & Value\\
\hline
log \oiii/\Hb & 0.3867 \\
log \nii/\Ha & 0.1498 \\
log \oii/\Hb & 0.717 \\
log \sii/\Ha & 0.0607 \\
log \oi/\Ha & -0.674 \\
\hline
\end{tabular}
\caption{Emission line ratios derived from the nuclear spectrum of CGCG~292$-$057.}
\end{center}
\end{table}

The SDSS data gives the observed stellar velocity dispersion, $\sigma_{*}$, which can be used
to estimate the black hole mass through the relation given by Tremaine et al. (2002). For CGCG~292$-$057, 
the velocity dispersion of 242.6 $\pm$ 13.6 km\,s$^{-1}$ 
gives a mass for the central black hole (or black hole pair) of 
log(M$_{\rm BH}$/M$_{\odot}$) = 8.47 $\pm$ 0.32. However, since the relation is 
valid for bulge-dominated systems the black hole mass may be underestimated for
such a highly disturbed galaxy. 



\begin{figure*}
\centering
\includegraphics[width=18cm,angle=0]{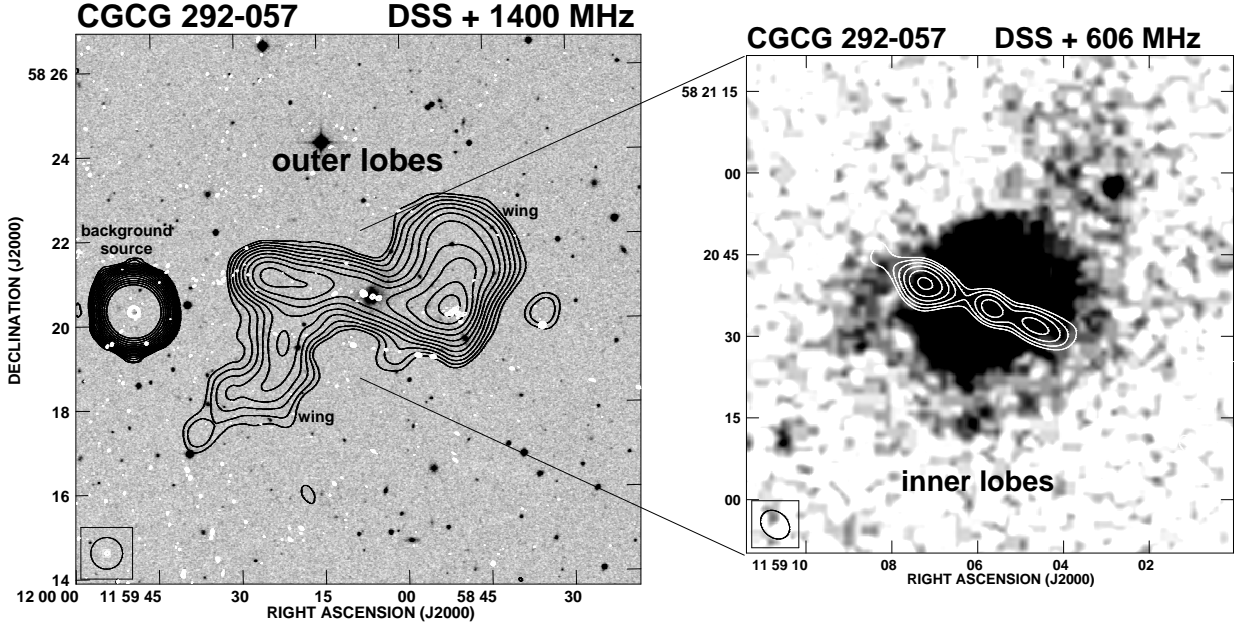}
\caption{1400~MHz VLA images of CGCG~292$-$057. (left panel) Contour maps of the 
entire source from the NVSS (black) and FIRST (white) surveys overlayed on the 
optical field from the DSS. The NVSS contour levels are spaced by factors 
of $\sqrt{2}$, and the first contour is at 1.35~mJy~beam$^{-1}$. (right panel) 
Contour map of the central part of the source from the GMRT 606~MHz map 
overlayed on DSS. The contour levels are spaced by factors of $\sqrt{2}$, 
and the first contour is at 1.5~mJy~beam$^{-1}$. The relative sizes of the beams are
indicated by the circles in the bottom left corner of each image.}
\label{fig:firstnvss}
\end{figure*}


\section{Multi-wavelength radio analysis}
\label{sec:radio}

\subsection{Interferometric maps}
\label{sec:RD}

CGCG~292$-$057 has been mapped in three northern-sky radio surveys: the NRAO VLA 
Sky Survey (NVSS; Condon et al. 1998), the Faint Images of the Radio Sky 
at Twenty-cm survey (FIRST; Becker, White \& Helfand 1995), and the Westerbork Northern 
Sky Survey (WENSS; Rengelink et al. 1997). 

The NVSS map is a `cube' of three planes containing the Stokes $I$, $Q$, and $U$ images. 
The r.m.s. is about 0.45 for Stokes $I$ and 0.29~mJy~beam$^{-1}$ for Stokes $Q$ and $U$.
The 1400~MHz map of $45 \times 45$~arcsec resolution is shown in Fig.~\ref{fig:firstnvss} 
(left panel) with a greyscale optical image from the Digitized Sky Survey (DSS) overlaid. 
The total flux of the extended source is 325.6$\pm$10.1 mJy\footnote{The common flux density scale of 
our measurements is that of Baars et al. (1977)}.

\begin{figure}
\centering
\includegraphics[width=8cm,angle=0]{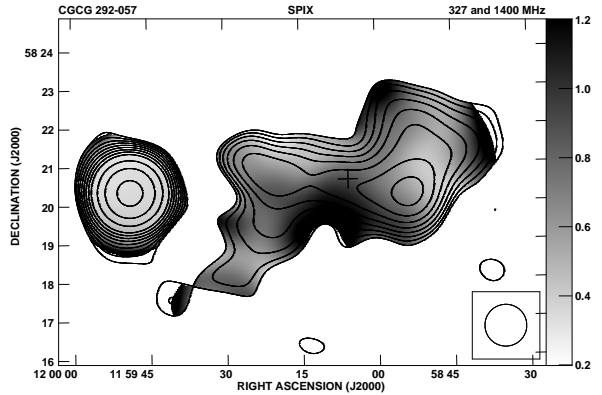}
\caption{The map of the spectral index $\alpha_{1400}^{327}$. Superimposed 
are 327~MHz total intensity contours from the WENSS survey, spaced by a factor 
of $\sqrt{2}$ and starting at 12~mJy~beam$^{-1}$. The size of the convolved beam of 
$65 \times 65$~arcsec\, is indicated by a circle in the bottom right 
corner of the image. The cross marks the central position of the host galaxy.}
\label{fig:spix}
\end{figure}

In addition to the structure described above, this radio source shows a second pair of coaxial lobes:
an inner pair of size of $\sim$ 23\,kpc, clearly visible at the high-resolution 
($5 \times 5$~arcsec) FIRST map. The flux densities of the three compact sources located 
at J2000.0 R.A. =  11$\rm ^h$ 59$\rm ^m$ 4\fs57, DEC = +58\degr 20$^\prime$ 33\farcs1; 
R.A. = 11$\rm ^h$ 59$\rm ^m$  5\fs67, Dec. = +58\degr 20$^\prime$ 35\farcs4; and 
R.A. = 11$\rm ^h$ 59$\rm ^m$ 7\fs28, Dec. = +58\degr 20$^\prime$ 40\farcs0 are respectively 2.14, 
2.11, and 4.36 mJy. The central component coincides well with the 
centre of the optical galaxy and the two side compact sources are located within 
the optical boundaries of the parent galaxy. 

The resolution of the 327~MHz WENSS 
image presented in Fig.~\ref{fig:spix} is about $54 \times 63$~arcsec and 
the r.m.s. noise across this field is 4.3~mJy beam$^{-1}$. The total integrated flux 
of this source is 851.3$\pm$32.2 mJy.

The central and eastern parts of the radio source CGCG~292$-$057 are apparent also in the VLA Low Frequency 
Sky Survey (VLSS; Cohen et al. 2007) at 74 MHz. 

We performed dedicated observations of this source with the Giant Metrewave Radio Telescope (GMRT) 
at 606~MHz. Details of the observations and data analysis will be published elsewhere 
(Jamrozy et al., in preparation), but the contour map of the inner part of this source is 
given in Fig. \ref{fig:firstnvss} (right panel). 

This source is also visible in the single-dish Green Bank 4850 MHz northern sky survey 
(GB6; Gregory et al. 1996), but because of the large beam size of this survey 
(230~arcsec~$\times$~196~arcmin) and a bright (1191.3$\pm$35.7 mJy at 1400 MHz) 
nearby background source (RA = 11$\rm ^h$ 59$\rm ^m$ 48\fs78, DEC = +58\degr 20$^\prime$ 20\farcs0) 
the view is strongly confused. Indeed, it appears that this nearby point-like source has caused 
confusion elsewhere in the literature. An object named CRATES J1159$+$5820 appears in the 
Combined Radio All-Sky Targeted Eight GHz Survey' (CRATES; Healey, et al. 2007). This survey 
presents a large sample of bright, compact, flat-spectrum sources; however, the authors mistakenly give the 
position of the extended radio galaxy (i.e. RA = 11$\rm ^h$ 59$\rm ^m$ 05\fs67, 
DEC = +58\degr 20$^\prime$ 35\farcs7), rather than the position of the nearby strong point source.
Therefore, and to avoid further confusion, in this paper we use the name of the parent galaxy,
CGCG~292$-$057, to refer to the extended radio source, rather than the name J1159$+$5820.

\subsection{Morphology}
\label{sec:RM}
At first glance, the extended ($\sim$5~arcmin) radio morphology of CGCG~292$-$057
resembles that of an FRII-type radio galaxy, although with some unusual features. 
It shows an X-like (Z-symmetric) structure characterized by two low-surface-brightness 
`wings'; oriented at an angle to the high-surface-brightness lobes. The wings and lobes pass symmetrically 
through the location of the host galaxy. 
The extended morphology of this source is reminiscent of NGC~326 (Murgia et al. 2001). It is possible 
that the structure of the wings and the outer lobes arises from similar processes to those in more typical X-shaped sources.
In addition, this source show second pair of coaxial lobes. Besides the outer one described above,
there is also an inner pair, well visible in the GMRT and FIRST maps. The outer lobes are believed 
to have formed during a previous cycle of jet activity, while the inner coaxial lobes probably due 
to some more recent nuclear activity.

Any prominent hotspots in the outer high-surface-brightness lobes were not detected in the high-resolution maps. However, 
there appears to be an elongated, diffuse shell-like structure at the position of a presumed 
hotspot in the western outer lobe. 

The radio power of the outer pair is 
$P_{\rm1400MHz} \simeq 2 \times 10^{24}{\rm W\,Hz^{-1}}$ which places the source in the FRII/FRI transition 
category. Merritt \& Ekers (2002) suggested that the radio source type may depend on the time scale for jet axis 
reorientation. Slow reorientation leads to an `S-shaped' FRI source, while rapid reorientation would
produce an intermediate-luminosity X-shaped source with a radio power near the FRII/FRI break. High-luminosity
FRII sources appeared when the jet reorientation occurred long time ago and the jets and `active' lobes are well 
aligned. Since CGCG~292$-$057 is of mixed type, and the wings are still easily visible it can be concluded that 
a rapid jet axis reorientation may have occurred in the not too distant past ($<10^{8}$ yr).

\begin{figure}
\centering
\includegraphics[width=8cm,angle=0]{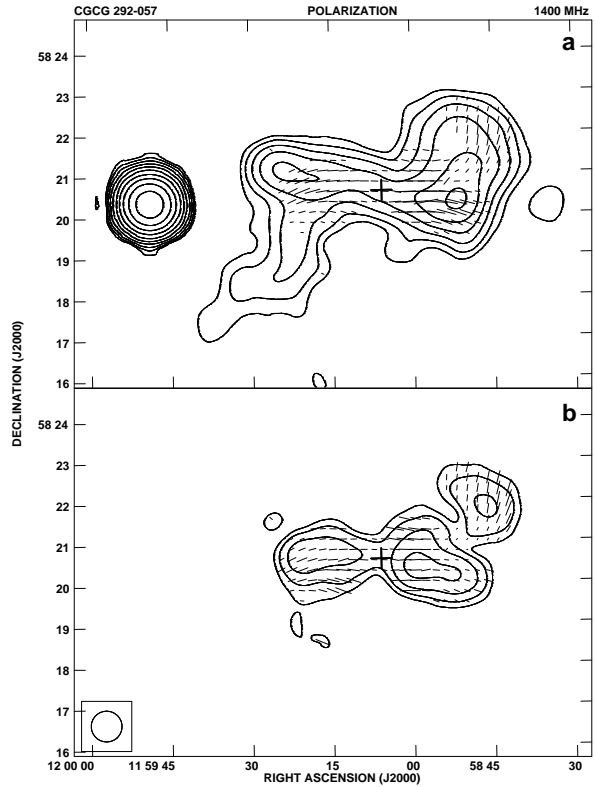}
\caption{1400 MHz NVSS polarimetric images. (a) Total intensity contours 
spaced by a factor of 2 starting at 1.35~mJy~beam$^{-1}$. Superimposed are 
\textbf{\textit{E}}-vectors with their lengths proportional to the polarized intensity, 
where 10~arcsec, corresponds to 2.5~mJy~beam$^{-1}$. (b) Linearly polarized 
intensity contours spaced by a factor of 2 starting with 0.9 mJy~beam$^{-1}$ with 
the vectors of the fractional linear polarization superimposed. A length of 
10~arcsec, corresponds to 25~per~cent of the fractional linear polarization. The 
cross marks the central position of the host galaxy.}
\label{fig:pol}
\end{figure}
  
The radio contours of the outer lobes (NVSS map) as well as of the inner lobes (GMRT map) 
overlayed on the DSS optical image are shown on Fig. \ref{fig:firstnvss}. To produce this figure the radio 
and optical maps were brought to a common scale using the \textsc{aips}\footnote{http://www.aips.nrao.edu} 
task \textsc{HGEOM}. We assumed that the astrometry of the NVSS, FIRST and DSS maps is unambiguously 
correct. The astrometry of the GMRT map was checked by comparing the positions of several point-like 
sources in the vicinity of 0\fdg5 around CGCG~292$-$057 as well as peaks of radio emission of the inner 
lobes and the radio core with those at the FIRST map which was taken as a reference. Small corrections 
to the GMRT map were incorporated.

Taking into account the morphological asymmetry of the inner lobes we estimate the inclination of the structure to
be about 79\degr, whereas the host galaxy is positioned almost ‘face-on’, as noted earlier. Several authors (e.g. Palimaka et al. 1979;
Guthrie 1980; Andreasyan \& Sol 1999) found a preference for the
orientation of the jet axis in powerful radio galaxies along the host minor
axis, and the tendency for this alignment is stronger for larger
radio sources. On the other hand, Saripalli \& Subrahmanyan (2009) noted that
X-shaped radio sources have radio axes within 50\degr of their host major
axes.

\subsection{Spectral index and polarization}
\label{sec:RSI}
We used the NVSS and WENSS images to create a spectral index map of CGCG~292$-$057.
First, the higher-resolution map was convolved to the resolution of the 327 MHz map.
Since misalignment of the total-power map at two frequencies could produce 
systematic errors in the spectral-index map, we co-registered the positions of several bright point-like field sources 
which surround the target source on both maps. Further, both maps were brought to a common scale using the 
\textsc{aips} task \textsc{HGEOM}. Finally, the spectral index map was obtained
using the \textsc{aips} task \textsc{COMB}. Regions with flux density values below 3$\times$r.m.s. 
were consider to be unreliable and blanked. The final grayscale map of the spectral index is shown in 
Fig. \ref{fig:spix}. 
We find that in the outer lobes there is a monotonic flattening of the radio spectrum 
from the center towards the presumed hotspot regions. This is a classical spectral signature seen 
in almost all normal FRII radio galaxies whose lobes are filled with backflow
material, which is older in the centre than on the edge of the radio galaxy.
The spectral index near the position of the host galaxy is $\alpha_{1400}^{327}\approx 0.9$
 and near the presumed hotspot $\alpha_{1400}^{327}\approx  0.6$ (S$\sim\nu^{-\alpha}$). 
The mean spectral index over the whole structure is about 0.76, and the mean spectral 
index over the south-eastern and north-western low surface-brightness wings 
is about 0.85 and 0.79, respectively. High values of the spectral index 
of the order of 1.3 at some points on the outskirts of this source might 
be due to missing flux in the 1400~MHz interferometric map.  
The lack of clearly steeper spectra and hence an older electron population in the wings 
is at first sight somewhat surprising. However, Dennet-Thorpe et al. (2002) and Lal \& Rao (2007) 
showed that the wings of a significant fraction of X-shaped sources have comparable or even flatter
spectral indices than found in the currently-active lobes. This contrasts with the expectations of 
models in which the wings of X-shaped sources are formed hydrodynamically or by jet reorientation, 
but could be in agreement with the Lal \& Rao (2005, 2007) hypothesis that such sources might be powered 
by a pair of associated and unresolved AGNs.

Maps of the linearly-polarized intensity and fractional polarization were made by combining the NVSS $Q$ and 
$U$ maps  with the \textsc{aips} task \textsc{COMB}. (We assume that the alignment of these maps is correct 
since we took them directly from the NVSS database as a cube of three aligned planes containing the Stokes $I$, $Q$, and $U$ images.)
These allow determination of the polarized flux density, the fractional polarization, 
and the polarization angle of the \textbf{\textit{E}}-vector. The total intensity NVSS map 
with the electric field \textbf{\textit{E}}-vectors superimposed is shown in Fig.~\ref{fig:pol}a. 
Fig.~\ref{fig:pol}b shows the linearly polarized intensity map with the 
vectors of fractional linear polarization superimposed. It can be seen 
that the whole structure apart of the south-eastern wing is strongly polarized. 
The total integrated polarized flux intensity of this source is 61 mJy, which 
gives $\sim$19~per~cent for the mean fractional polarization. Such a high value of the
fractional polarization is typical of X-shaped sources (Dennett-Thorpe 
et al. 2002). 



\section{Concluding Remarks}

CGCG~292$-$057 is the only radio galaxy which shows both double-double and X-like morphology and which has
also been identified with a merger galaxy. The total flux of the entire source at 
1400~MHz corresponds to a monochromatic radio power of 
$P_{\rm1400MHz} \simeq 2 \times 10^{24}{\rm W\,Hz^{-1}}$. The outer lobes 
are highly polarized (almost 20~per~cent at 1400 MHz), as is typical of X-shaped radio 
sources. The Mt. Suhora and the LCOGT optical observations show fine structure with tidal features and tails 
confirming that the parent is a merger galaxy.
The host, CGCG~292$-$057, is a LINER-type galaxy with a black hole of relatively 
low mass, $\log(M_{\rm BH}/M_{\odot}) = 8.47$, and possibly with double-peaked narrow emission 
lines. 

These features make this galaxy a primary test target for models such as those presented by Liu (2004), 
in which X-shaped structures may be due to a realignment of the SMBBH interacting with the accretion disc, or by Liu et al. (2003), who presented a scenario to explain the interruption and restarting of the jet formation in DDRGs. In the latter case, after a minor galaxy merger the secondary black hole migrates inwards, disrupting the inner parts of the accretion disc. The gap in the accretion disc expands after the binary black hole coalesces, leading to an interruption in jet formation. Jet activity restarts following the inflow of new material into the central region.

The uniqueness of CGCG~292$-$057 arises from the apparently short time-scale of the galactic merger, subsequent jet reorientation and restarted activity, which allows us to simultaneously observe the results of these processes -- thus making it an ideal laboratory for the investigation of the evolution of radio galaxies.


\section*{ACKNOWLEDGMENTS}

We acknowledge the LCOGT DDT allocation for this project and Melissa Graham for
help in scheduling observations. We thank the anonymous referee for helpful comments 
which improved the paper.
This project was supported in part by MNiSW funds for scientific research
in the years 2009-2012 under agreement no. 3812/B/H03/2009/36.

{}

\end{document}